\documentclass{jpsj2}
\usepackage{graphicx}

\title{Superconductivity in Multilayer Perovskite: Weak Coupling Analysis}

\author{Shigeru Koikegami
\footnote{E-mail address : shigeru.koikegami@aist.go.jp} 
and Takashi Yanagisawa}

\inst{Nanoelectronics Research Institute, 
AIST Tsukuba Central 2, Tsukuba 305-8568, Japan}

\recdate{December 3, 2005}

\abst{We investigate the superconductivity of a three-dimensional 
d-p model with a multilayer perovskite structure 
on the basis of the second-order perturabation theory within the weak 
coupling framework. Our model has been designed with multilayer 
high-$T_{\mathrm{c}}$ superconducting cuprates in mind. In our 
model, multiple Fermi surfaces appear, and the component of 
a superconducting gap function develops on each band. 
We have found that the multilayer structure can stabilize 
the superconductivity in a wide doping range.}

\kword{superconductivity, three-dimensional d-p model, multilayer perovskite,
second-order perturbation theory}

\begin{document}
\sloppy
\maketitle

\section{Introduction}
In the last decade, 
high-$T_{\mathrm{c}}$ superconducting cuprates 
(HTSCs) with multilayer structure 
have been investigated extensively by various experimental 
methods~\cite{Ihara1997,Watanabe2000,Ihara2000}. 
Multilayer cuprates typically have higher $T_{\mathrm{c}}$ 
than single-layer ones. The nuclear magnetic resonance (NMR) 
measurement systematically reveals the local hole concentration on 
each layer of the multilayer HTSCs. 
In previous excellent studies~\cite{Tokunaga2000,Kotegawa2001}, each layer 
has been crystallographically classified into an outer or 
inner CuO$_2$ plane in a unit cell. 
The authors found the relationship between the difference in 
the local hole concentration among these two types of planes 
and $T_{\mathrm{c}}$, and they considered the condition in which 
$T_{\mathrm{c}}$ can be maximized. The results suggest that 
multiple Fermi surfaces are involved in superconductivity, 
and that an extensive theory on multilayer materials 
should be constructed on the basis of the model with multibands. 

The above NMR study mainly revealed the characteristic 
of the multilayer compounds 
for $n \geq 3$, where $n$ is the number of CuO$_2$ planes per unit cell. 
The characteristic feature of 
multiband systems can also appear explicitly in a bilayer compound. 
In Bi$_2$Sr$_2$CaCu$_2$O$_{8+\delta}$ (Bi2212), 
which is another typical bilayer material, 
the high-resolution angle-resolved-photoemission 
spectroscopy (ARPES) successfully revealed the doubling
of a band near the Fermi level~\cite{Chuang2001,Feng2001}. 
This splitting is negligible 
along the $(0,0)\rightarrow(\pi,\pi)$ nodal line 
and maximum at $(\pi,0)$ in momentum space. This momentum dependence of 
energy splitting is qualitatively consistent with the LDA prediction for 
YBa$_2$Cu$_3$O$_7$~\cite{Andersen1995}, which is another 
bilayer material. 
This LDA calculation predicted that a Cu $4s$ orbital 
has a transfer integral between that in the other layer. This 
interlayer coupling causes a single band to 
split into antibonding and bonding bands. 

A multilayer model has been theoretically studied since the early 1990s 
by fluctuation exchange (FLEX) approximation~\cite{Bulut1992} 
and by quantum Monte Carlo (QMC) simulation~\cite{Bulut1992,Scalettar1994} 
on the basis of the two-dimensional (2D) multilayer Hubbard model. 
These studies have shown that the $d_{x^2-y^2}$-wave 
pairing correlations are reduced by interlayer transfer, which 
is independent of the in-plane momenta, $k_x$ and $k_y$. However, 
when we introduce interlayer transfer into our model, 
it is important to consider the symmetry of interlayer 
transfer integrals in which Cu $4s$ electrons participate. 
Liechtenstein \textit{et al.} considered this point 
and studied the extended Hubbard model with anisotropic 
interlayer hopping, 
using the FLEX approximation~\cite{Liechtenstein1996}. Although they 
could not reproduce the experimental result of bilayer materials 
having higher $T_{\mathrm c}$ than single-layer ones, their work should be 
appreciated as the first theoretical analysis of the superconductivity of 
realistic multilayer systems. 
All the theoretical works have been done on the basis of the 2D model 
Hamiltonian. We feel that we should investigate 
the {\textit{three-dimensional}} model Hamiltonian with anisotropic 
interlayer hopping to estimate $T_{\mathrm c}$ 
of multilayer materials. 

In this study, we investigate the superconductivity 
of multilayer perovskite. 
We adopt the three-dimensional (3D) d-p model with 
anisotropic interlayer transfers as our model Hamiltonian.  
In our model, we introduce such a small on-site 
Coulomb interaction that the second-order perturbation theory (SOPT) 
can be justified. We can treat the superconductivity within the 
weak coupling analysis because, in our model, the effective interaction for 
Cooper pairing is so small that only the electrons on the Fermi surface 
are involved in the superconductivity. The weak coupling formalism for the 
repulsive interaction model, 
since the pioneering work by Kohn and Luttinger~\cite{Kohn1965},
has been developed by many theorists
~\cite{Fay1968,Nakajima1973,Anderson1973,Kagan1988,Hlubina1999}. 
Recently, this formalism was applied, by Kondo, to 
the 2D Hubbard model with the formulation applicable even for the 
case with a very small effective interaction~\cite{JKondo2001}. 
We apply Kondo's formulation to our model with multiple Fermi surfaces, 
and clarify how the superconducting gap depends on $n$ of layers. 
We can show that the calculation on the basis of 3D model 
Hamiltonian is requisite for the true estimation of $T_{\mathrm c}$ of 
multilayer materials. Our obtained results are not only to be compared 
with the actual $T_{\mathrm c}$ of multilayer materials but to be considered 
as a guide for designing materials with high $T_{\mathrm c}$.

\section{Formulation}
\begin{figure}
\includegraphics[width=8.5cm]{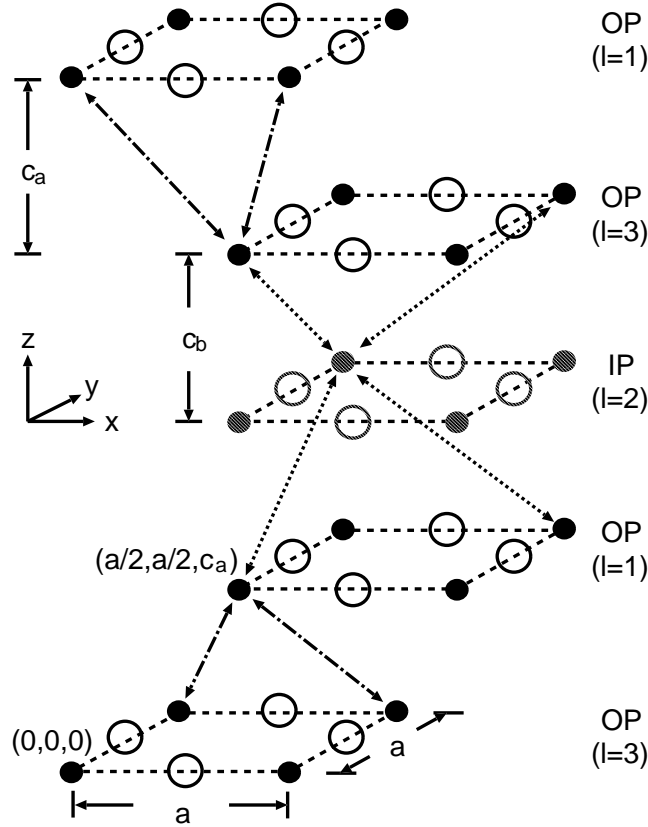}
\caption{\label{figure:1}Crystal structure modeled after trilayered perovskite. Filled (Open) circles represent copper (oxygen) cites. Dotted (Dot-dashed) arrows conecting copper cites correspond to $\zeta_{\mib{k}}^d$ ($\zeta_{\mib{k}}^z$) in the text.}
\end{figure}

We can decompose our 3D d-p 
model with the $n$-layer perovskite structure into several parts as follows:
\begin{align}
H  = \sum_{l=1}^n & \left[\right.
H_l^0+H_{l+1,\,l}^0+H_l^\prime \nonumber \\
    & -\mu \sum_{\mib{k} \sigma}
(d_{\mib{k}\,l \sigma}^\dagger d_{\mib{k}\,l \sigma}
+p_{\mib{k}\,l \sigma}^{x \dagger} p_{\mib{k}\,l \sigma}^x
+p_{\mib{k}\,l \sigma}^{y \dagger} p_{\mib{k}\,l \sigma}^y)
\left.\right], 
\nonumber \\
\label{eq:11}
\end{align}
where $d_{\mib{k}\,l \sigma}$ $(d_{\mib{k}\,l \sigma}^{\dagger})$, 
$p_{\mib{k}\,l \sigma}^x$ $(p_{\mib{k}\,l \sigma}^{x \dagger})$ and
$p_{\mib{k}\,l \sigma}^y$ $(p_{\mib{k}\,l \sigma}^{y \dagger})$ are
the annihilation (creation) operators for d-, p$^x$- and p$^y$-electrons of
momentum $\mib{k}$ and spin $\sigma=\{\uparrow,\downarrow\}$ 
on the $l$-th layer, respectively. We define $n+1 \equiv 1$, 
and the chemical potential is represented by $\mu$. 
The noninteracting parts in eq.~(\ref{eq:11}), 
i.e., $H_l^0$ and $H_{l+1,\,l}^0$ , are represented by
\begin{align}
H_l^0 & \nonumber \\
 = & \sum_{\mib{k} \sigma} 
\left(d_{\mib{k}\,l \sigma}^\dagger\,
p_{\mib{k}\,l \sigma}^{x \dagger}\,
p_{\mib{k}\,l \sigma}^{y \dagger}\right)\hspace{-0.3em} 
\left(
\begin{array}{ccc} \Delta_l & 
\zeta_{\mib{k}}^x  & \zeta_{\mib{k}}^y \\ 
-\zeta_{\mib{k}}^x & 0 & \zeta_{\mib{k}}^p \\
-\zeta_{\mib{k}}^y & \zeta_{\mib{k}}^p & 0 \\
\end{array} \right)\hspace{-0.5em}
\left( \begin{array}{c} d_{\mib{k}\,l \sigma}
\\ p_{\mib{k}\,l \sigma}^x 
\\ p_{\mib{k}\,l \sigma}^y \\
\end{array} \right) \nonumber \\
\label{eq:8}
\end{align}
and
\begin{equation}
H_{l+1,\,l}^0 
= \left\{
\begin{array}{ll}
\sum_{\mib{k} \sigma}\zeta_{\mib{k}}^z\,
d_{\mib{k}\,l+1 \sigma}^\dagger d_{\mib{k}\,l \sigma}+{\mathrm{h.c.}} 
& l=n \\
\sum_{\mib{k} \sigma}\zeta_{\mib{k}}^d\,
d_{\mib{k}\,l+1 \sigma}^\dagger d_{\mib{k}\,l \sigma}+{\mathrm{h.c.}} 
& l<n
\end{array}
\right..
\label{eq:9}
\end{equation}
In eqs.~(\ref{eq:8}) and (\ref{eq:9}) 
we take the lattice constant of the square
lattice formed of Cu sites, $a$, as the unit length. Using it, 
we can represent
$\zeta_{\mib{k}}^p=4t_{pp} \sin \frac{k_x}{2} \sin \frac{k_y}{2}$, 
$\zeta_{\mib{k}}^x=2{\rm i}t_{dp} \sin \frac{k_x}{2}$,
$\zeta_{\mib{k}}^y=2{\rm i}t_{dp} \sin \frac{k_y}{2}$,
$\zeta_{\mib{k}}^z=t_\perp \cos \frac{k_x}{2} \cos \frac{k_y}{2}e^{{\mathrm i}k_zr_a}$, and 
$\zeta_{\mib{k}}^d=\frac{t_{dd}}{4} (\cos k_x-\cos k_y)^2e^{{\mathrm i}k_zr_b}$, where $r_a=\frac{c_a}{c_a+(n-1)c_b}$ and 
$r_b=\frac{c_b}{c_a+(n-1)c_b}$. $c_a$ and $c_b$ represent 
the distances among CuO$_2$ planes, as shown in Fig.~\ref{figure:1}.
Hereafter, we use 'IP' and 'OP' as 
abbreviations for inner CuO$_2$ planes ($1<l<n$) and outer CuO$_2$ planes 
($l=1 \mbox{ or } n$), respectively. 
$\Delta_l$ is the hybridization gap energy on the $l$-th layer 
between d- and p-orbitals. When $n \geq 3$, our model has several 
crystallographically inhomogeneous CuO$_2$ planes, and 
$\Delta_l$ varies according to $l$. We can define $\Delta_l$ as
\begin{equation}
\Delta_l = \left\{
\begin{array}{ll}
\Delta_{dp} & l=1 \mbox{ or } n \\
\Delta_{dp}-\frac{\alpha}{r_a} & 1<l<n
\end{array}
\right.,
\label{eq:14}
\end{equation}
where $\alpha$ is the constant chosen so as to reproduce 
the difference in doped carriers between OP and IP. 
We consider only the on-site Coulomb repulsion among d-electrons. 
Thus, the interacting part $H_l^\prime$ in eq.~(\ref{eq:11}) is described as
\begin{equation}
H_l^\prime = \frac{U}{N} \sum_{\mib{k} \mib{k}^\prime \mib{q}}
        d_{\mib{k}+\mib{q}\,l \uparrow}^{\dagger} 
	d_{\mib{k}^\prime-\mib{q}\,l \downarrow}^{\dagger} 
	d_{\mib{k}^\prime\,l \downarrow} 
	d_{\mib{k}\,l \uparrow}.
\label{eq:10}
\end{equation}
In eq.~(\ref{eq:10}), 
$N$ is the number of $\mib{k}$-space lattice points in the first
Brillouin zone (FBZ), which is equal to the number of Cu sites in the
real space. 

In the following part, we assume that only the electrons on the Fermi surface 
of the same band can have singlet pair instability. For our $n$-layer model, 
$n$ d-like bands always intersect with the Fermi level. Thus, according to 
the BCS theory, we have the following self-consistent equation for the pair 
function on the $\lambda$-th d-like band, $\Phi_{\mib{k}\lambda}$:
\begin{equation}
\Phi_{\mib{k}\lambda}=
-\frac{1}{2N}\!\sum_{ij\mib{k}^\prime \nu}%
V_{ij}(\mib{k}+\mib{k}^\prime)
\frac{z_{i \lambda}(\mib{k})z_{j \nu}(\mib{k}^\prime)}
{\sqrt{\left(\varepsilon_{\mib{k}^\prime \nu}-\mu \right)^2+
\left(\Phi_{\mib{k}^\prime \nu}\right)^2}}\,\Phi_{\mib{k}^\prime \nu},
\label{eq:3}
\end{equation}
where $i,j=1,\ldots,n$ (layer indices) and $\lambda,\nu=1,\ldots,n$ 
(d-like band indices). $V_{ij}(\mib{q})$ represents the effective 
singlet pair scattering between a d-electron on the $i$-th layer and one 
on the $j$-th layer. $\varepsilon_{\mib{k}\nu}$ represents the energy 
dispersion of the $\nu$-th d-like band, and $z_{i \lambda}(\mib{k})$ 
represents the matrix element of unitary transformation. 
They are obtained by solving the eigenequation 
for the noninteracting part in eq.~(\ref{eq:11}).
We set $\Phi_{\mib{k} \lambda}=
\Delta_{\mathrm{sc}}\!\cdot\!\Psi_{\mib{k} \lambda}$, 
where $\Delta_{\mathrm{sc}}$ denotes the magnitude of $\Phi_{\mib{k} \lambda}$ 
and $\Psi_{\mib{k} \lambda}$ represents 
its $\mib{k}$-dependence on the $\lambda$-th d-like band. 
On the basis of Kondo's argument,~\cite{JKondo2001} 
retaining only the divergent term, we can rewrite eq.~(\ref{eq:3}) as 
\begin{equation}
\begin{split}
\Psi_{\mib{k}\lambda} &= \\
& \hspace{-2em} \log_e \Delta_{\mathrm{sc}}\cdot \frac{1}{N}
\sum_{ij\mib{k}^\prime \nu}V_{ij}(\mib{k}+\mib{k}^\prime)
 z_{i \lambda}(\mib{k})z_{j \nu}(\mib{k}^\prime)
\delta(\varepsilon_{\mib{k}^\prime \nu}-\mu)
\Psi_{\mib{k}^\prime \nu}
\label{eq:4}
\end{split}
\end{equation}
for very small $\Delta_{\mathrm{sc}}$. Equation (\ref{eq:4}) 
is a homogeneous integral equation for 
$\Psi_{\mib{k}\lambda}$ with the eigenvalue of $1/\log \Delta_{\mathrm{sc}}$. 
We are interested in obtaining the most stable superconducting state, 
thus we must find the eigenvector $\Psi_{\mib{k}\lambda}$ with the smallest
eigenvalue $1/\log \Delta_{\mathrm{sc}}$ using eq.~(\ref{eq:4}) 
when $\Delta_{\mathrm{sc}}$ is maximum. In our previous paper, 
we confirmed that 
the most stable pairing state near half-filling is the $d_{x^2-y^2}$-wave, by 
a similar approach based on 2D d-p model~\cite{Koikegami2001}. Hence, 
when we assume that 
\begin{equation}
\Psi_{\mib{k}\lambda} = a_\lambda(k_z)(\cos k_x-\cos k_y), 
\label{eq:12}
\end{equation}
we can safely reduce our original eigenvalue problem 
for $\Psi_{\mib{k}\lambda}$ to an eigenvalue problem for $a_\lambda(k_z)$ 
in order to seek only the most stable pairing state. Furthermore, considering 
the symmetry of $\Psi_{\mib{k}\lambda}$ in eq.~(\ref{eq:12}), we can 
take
\begin{equation}
\begin{split}
V_{ij}(\mib{q}) & = U^2\chi_{ij}(\mib{q}) \\ 
& = \frac{U^2}{N}\sum_{\mib{k}\xi \eta}
z_{i\xi}(\mib{q}+\mib{k})z_{j\eta}(\mib{k})
\frac{\left(1-f_{\mib{q}+\mib{k}\xi}\right)f_{\mib{k}\eta}}
{\varepsilon_{\mib{q}+\mib{k}\xi}-\varepsilon_{\mib{k}\eta}}
\label{eq:1}
\end{split}
\end{equation}
within SOPT. In eq.~(\ref{eq:1}), 
\begin{equation}
f_{\mib{k}\eta} = \frac{1}{2}
\left[1-\tanh\left(\frac{\varepsilon_{\mib{k}\eta}-\mu}{2T}
\right)\right],
\label{eq:7}
\end{equation} 
and $T$ denotes the temperature. 

\section{Results and Discussion}

In our present analyses, all $\varepsilon_{\mib{k}\nu}$ and 
$z_{i \lambda}(\mib{k})$ in eq.~(\ref{eq:4}) 
are first calculated for $64^3$ $\mib{k}-$points 
on an equally spaced mesh in FBZ for each band. Then, we calculate 
$V_{ij}(\mib{k}+\mib{k}^\prime)$ in eq.~(\ref{eq:4}) only for $\mib{k}-$ and 
$\mib{k}^\prime-$points satisfying the conditions 
$\varepsilon_{\mib{k} \lambda}=\mu$ and 
$\varepsilon_{\mib{k}^\prime \nu}=\mu$, respectively. 
When we calculate $V_{ij}(\mib{k}+\mib{k}^\prime)$ according to 
eqs.~(\ref{eq:1}) and (\ref{eq:7}), 
we set the temperature $T=0.001\,$eV$\sim 10\,$K, at which our 
system can be considered to behave similarly to that in the ground state. 
These calculations have been performed at $U=0.5\,$eV, 
where magnetic instabilities cannot occur. We take $c_a=0.3$ and $c_b=0.7$ 
for all $n$. Other common parameters are summarized in Table~\ref{table:1}. 
\begin{table}[h]
\begin{tabular}{ccccc}
$t_{dp}$ & $t_{pp}$ & $t_{dd}$ & $t_\perp$ & $\Delta_{dp}$ \\ \hline
  $1.00\,$eV & $-300\,$meV & $10\,$meV & $5\,$meV & $2.10\,$eV \\
\end{tabular}
\caption{\label{table:1}Transfer and hybridization gap energies.}
\end{table}
In order to solve eq.~(\ref{eq:4}) practically, 
we substitute $\Psi$ into both sides of eq.~(\ref{eq:4}) 
using eq.~(\ref{eq:12}) 
and integrate for $k_x$, $k_x^\prime$, $k_y$, and $k_y^\prime$. 
Thus, we reduce 
eq.~(\ref{eq:4}) to the eigenequation for $a_\lambda(k_z)$. 
When we solve it numerically by the standard 
method, we can finally obtain both the eigenvalue, 
$1/\log_e \Delta_{\mathrm{sc}}$, and 
the eigenvector, $a_\lambda(k_z)$. 

\begin{figure}
\includegraphics[width=8.3cm]{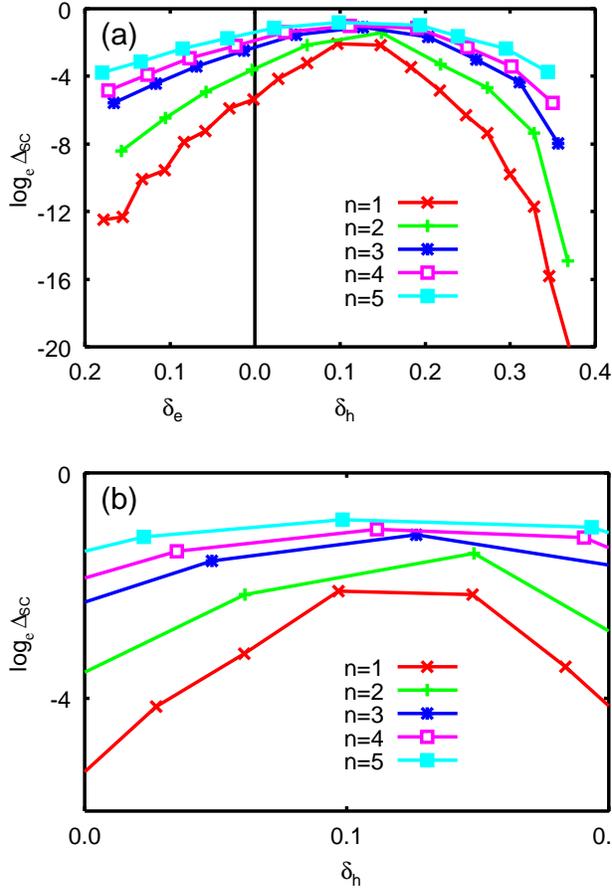}
\caption{\label{figure:5}(a) $\log \Delta_{\mathrm{sc}}$ vs 
$\delta_{\mathrm h}$ (hole-doped) or $\delta_{\mathrm e}$ (electron-doped) 
in the cases of $n=1,2,3,4,$ and $5$. (b) Magnification of (a) 
in the region of $0.0 \leq \delta_{\mathrm h} \leq 0.2$.}
\end{figure}

First, we summarize our results on $\Delta_{\mathrm{sc}}$ vs 
$\delta_{\mathrm h}$($\delta_{\mathrm e}$) in Fig.~\ref{figure:5}. 
As discussed in our previous paper on the 
two-dimensional (2D) d-p model~\cite{Koikegami2001}, 
the existence of a Van Hove singularity (VHS) 
at the Fermi level causes a high density of states (DOS) and enhances 
$\Delta_{\mathrm{sc}}$. In the 2D d-p model the parts of the Fermi 
surface with VHS are distributed as {\textit{lines}}. 
Therefore $\Delta_{\mathrm{sc}}$ 
varies drastically in the neighborhood of the doping point at which the 
Fermi surface has VHS. This is in contrast with the case in the 3D d-p model. 
Although the energy dispersion along the $c$-axis introduced in our analyses 
is very weak, as indicated in Table~\ref{table:1}, 
the parts of the Fermi surface with VHS 
are distributed as {\textit{points}}. Thus, the transition of 
$\Delta_{\mathrm{sc}}$ in the 3D model is milder than that 
in the 2D model, as seen in Fig.~\ref{figure:5}. 

\begin{figure}
\includegraphics[width=8.3cm]{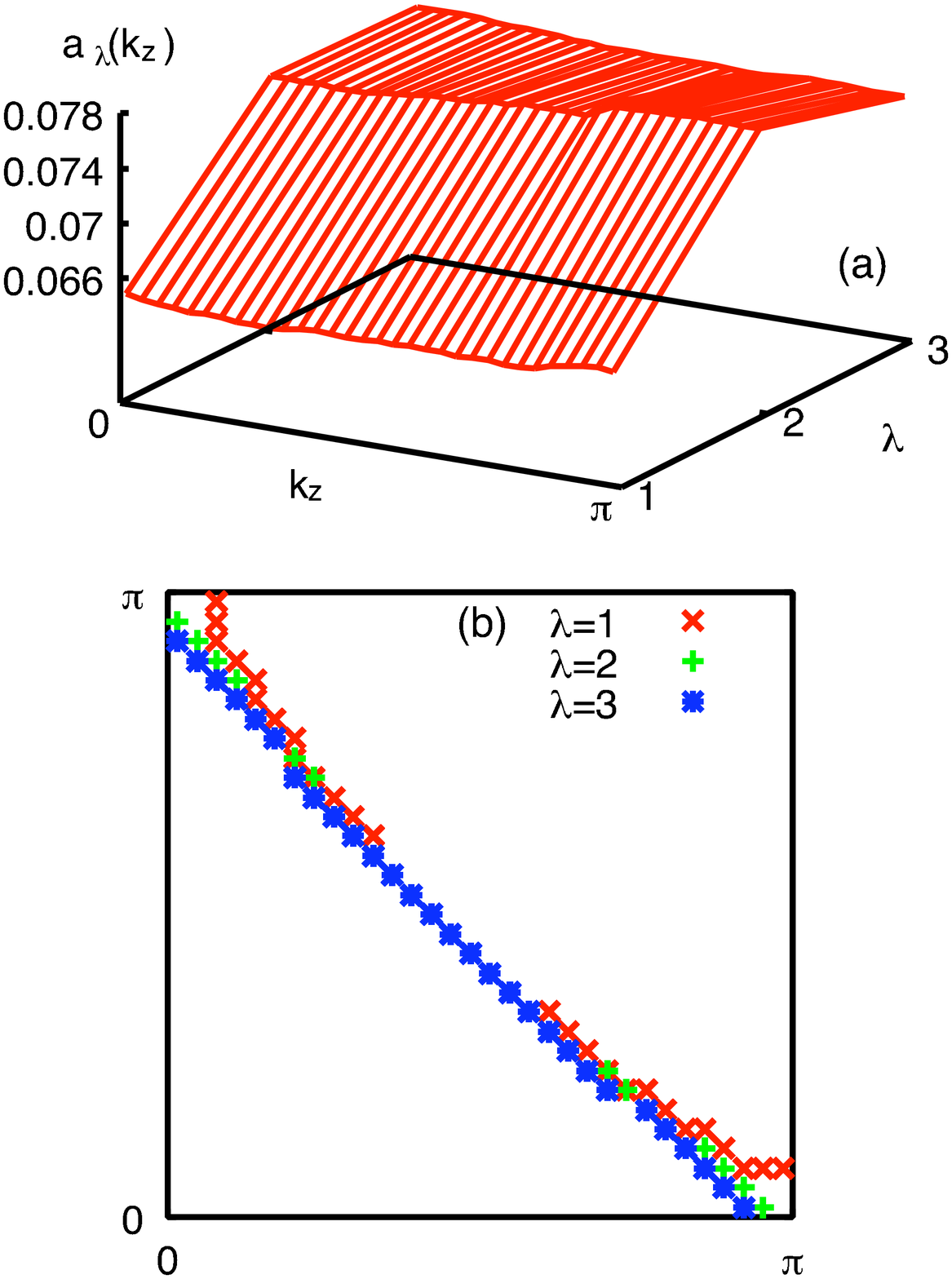}
\caption{\label{figure:2}The results for $n = 3$ and $\delta_{\mathrm{h}}=0.127$. (a) The eigensolution $a_\lambda(k_z)$. (b) The Fermi surface projected onto the plane with $k_z=0$.}
\end{figure}

\begin{figure}
\includegraphics[width=8.3cm]{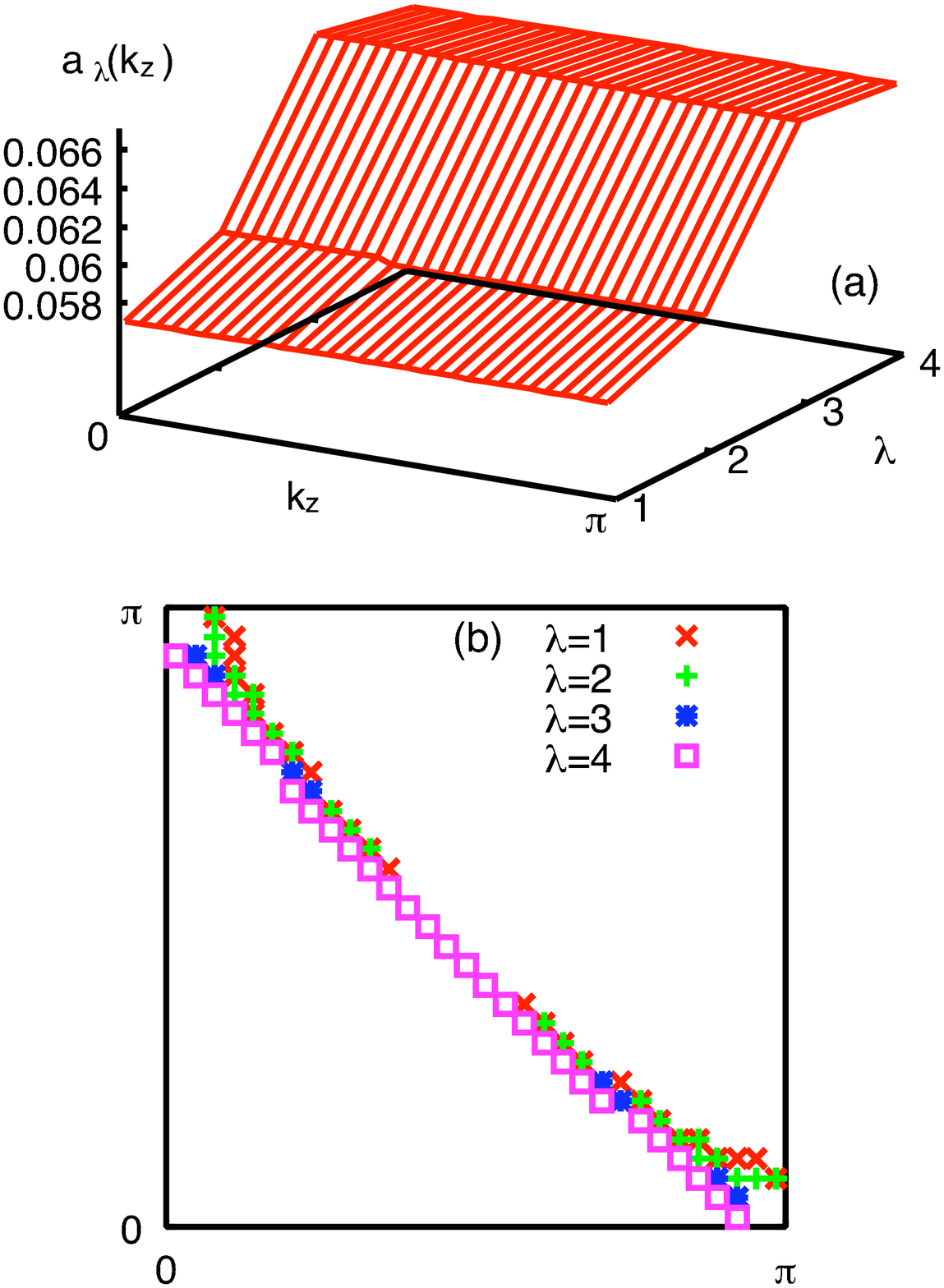}
\caption{\label{figure:3}The results for $n = 4$ and $\delta_{\mathrm{h}}=0.112$. (a) The eigensolution $a_\lambda(k_z)$. (b) The Fermi surface projected onto the plane with $k_z=0$.}
\end{figure}

\begin{figure}
\includegraphics[width=8.3cm]{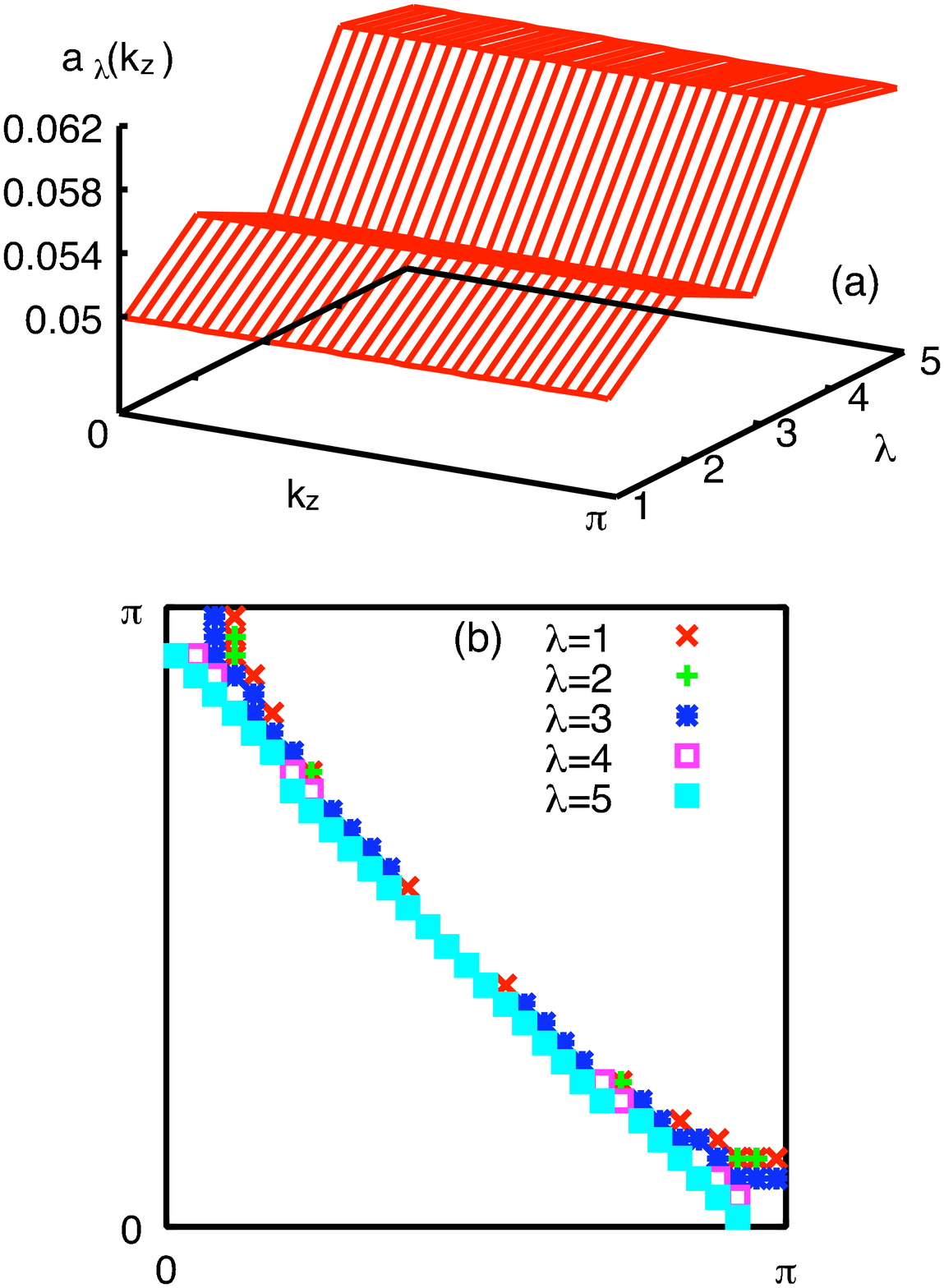}
\caption{\label{figure:4}The results for $n = 5$ and $\delta_{\mathrm{h}}=0.098$. (a) The eigensolution $a_\lambda(k_z)$. (b) The Fermi surface projected onto the plane with $k_z=0$.}
\end{figure}

Furthermore, in Fig.~\ref{figure:5} we can clearly recognize that 
the enhanced $\Delta_{\mathrm{sc}}$ prevails in a wider doping region 
with larger $n$. This result is caused by 
the {\textit{multilayering effect}} introduced as described by 
eq.~(\ref{eq:14}). 
In order to explain how the multilayering effect occurs, 
we show the eigensolutions, 
$a_\lambda(k_z)$, defined by eq.(\ref{eq:12}), 
and the Fermi surfaces for the cases with $n = 3,4,$ and $5$ 
in Figs.~\ref{figure:2},~\ref{figure:3}, and~\ref{figure:4}, respectively. 

In Fig.~\ref{figure:2} the highest amplitude of the eigensolution appears 
in the band with $\lambda = 2$, whose Fermi surface most closely 
approaches the VHS points, i.e., $(\pm\pi,0,[-\pi,\pi])$ and $(0,\pm\pi,[-\pi,\pi])$. 
Thus, the band with $\lambda = 2$ has the largest DOS near VHS points 
and is dominant in the superconductivity. 
For the same reason, the largest amplitude of the eigensolution appears in the band with $\lambda = 3$ and in the one with $\lambda = 4$, 
as shown in Figs.~\ref{figure:3} and \ref{figure:4}, respectively.

When we change the amount of doped carriers, the Fermi surface 
should be transformed. As a result, another band could then have 
the largest DOS near the VHS points and dominate the superconductivity. 
This possibility should be further increased more if our model has 
more alternative bands. Hence, 
the enhanced $\Delta_{\mathrm{sc}}$ tends to prevail in a wider doping region 
with larger $n$. This tendency should remain when $n$ becomes much larger, 
as long as the conventional Fermi surface can be defined. 
However, the largest value of $\Delta_{\mathrm{sc}}$ should be 
saturated toward the intrinsic value for $n \rightarrow \infty$.

Hereafter, we turn our attention to the maximum $T_{\mathrm{c}}$ 
of the $n$-layered materials, which would be proportional to 
the maximum $\Delta_{\mathrm{sc}}$ in our calculated results. 
In several real materials, the largest $T_{\mathrm{c}}$ is achieved 
when $n=3$ or $n=4$, and $T_{\mathrm{c}}$ is rather low when $n=5$~\cite{Tokunaga2000,Kotegawa2001}. For such materials 
our assumption that the well-defined Fermi surface exists 
might not be valid. For example, in the five-layered compound 
HgBa$_2$Ca$_4$Cu$_5$O$_y$, the inner CuO$_2$ 
planes turn out to be antiferromagnetic 
on account of the strong electronic correlation~\cite{Kotegawa2004}. 
Concerning the strong electronic correlation, 
other theoretical works 
based on the 2D multilayer t-J model have been extensively carried out 
by Mori et al.~\cite{MMori2002,MMori2003}. 
Their approach would be better for explaining the results for such compounds.  

Although our results on $T_{\mathrm{c}}$ are not consistent with 
those for several real materials, our conclusion on the multilayering effect 
is clearly applicable to other real materials. 
Indeed, (Cu,C)Ba$_2$Ca$_3$Cu$_4$O$_{12+y}$ (Cu1234), 
has a high $T_{\mathrm c}$ even though it is in the heavily overdoped 
region~\cite{Ihara1997,Ihara2000,Watanabe2000}. 
Cu1234 has been revealed, by NMR experiment, 
to have doped holes that are almost uniformly 
distributed into each layer~\cite{Tokunaga2000,Kotegawa2001}. Thus,  
our assumption concerning the Fermi surface is considered to 
be valid for Cu1234.

\section{Summary}
We demonstrated that 
the ground state of the 3D d-p model with a multilayer perovskite structure 
can be in the $d_{x^2-y^2}$-wave superconducting state up to 
the second-order in the perturbation theory framework. 
In the multilayer system, the region with large 
$\log \Delta_{\mathrm{sc}}$ can expand further. This is caused by  
the multilayering effect, which can increase 
the chance that the Fermi surface has VHS and 
can maintain a high DOS around the Fermi level over a wide doping region.  
This multilayering effect works very well when the unit cell contains 
more layers, as long as a well-defined Fermi surface exists. 

\section*{Acknowledgments}
The authors thank to Dr. Y. Aiura for providing his group's ARPES results and for invaluable discussions. 
The authors are also grateful to Professor J. Kondo and Professor K. Yamaji 
for their invaluable comments. 
S. K. also thanks Dr. M. Mori and Mr. S. Sasaki for stimulating discussions.

\end{document}